\documentclass[aps,prl,twocolumn,showpacs,superscriptaddress]{revtex4}
\usepackage{graphicx}
\begin{document}
%\preprint{}
\title{Spin excitations in optimally P-doped BaFe$_2$(As$_{0.7}$P$_{0.3}$)$_2$ superconductor}

\author{Ding Hu}
\affiliation{Center for Advanced Quantum Studies and Department of Physics, Beijing Normal University, Beijing 100875, China}
\affiliation{Beijing National Laboratory for Condensed Matter
Physics, Institute of Physics, Chinese Academy of Sciences, Beijing
100190, China}

\author{Zhiping Yin}
\email{yinzhiping@bnu.edu.cn}
\affiliation{Center for Advanced Quantum Studies and Department of Physics, Beijing Normal University, Beijing 100875, China}
\affiliation{Department of Physics, Rutgers University, Piscataway, NJ 08854, USA}

\author{Wenliang Zhang}
\affiliation{Beijing National Laboratory for Condensed Matter
Physics, Institute of Physics, Chinese Academy of Sciences, Beijing
100190, China}

\author{R. A. Ewings}
\affiliation{ISIS Facility, STFC Rutherford Appleton Laboratory, Harwell Oxford, Didcot, OX11 0QX, United Kingdom}

\author{Kazuhiko Ikeuchi}
\affiliation{Research Center for Neutron Science and Technology, Comprehensive Research Organization for Science and Society (CROSS), Tokai, Ibaraki 319-1106, Japan}

\author{Mitsutaka Nakamura}
\affiliation{Materials and Life Science Division, J-PARC Center, Tokai, Ibaraki 319-1195, Japan}

\author{Bertrand Roessli}
\affiliation{Laboratory for Neutron Scattering and Imaging, Paul Scherrer Institut, CH-5232 Villigen, Switzerland}

\author{Yuan Wei}
\affiliation{Beijing National Laboratory for Condensed Matter
Physics, Institute of Physics, Chinese Academy of Sciences, Beijing
100190, China}

\author{Lingxiao Zhao}
\affiliation{Beijing National Laboratory for Condensed Matter
Physics, Institute of Physics, Chinese Academy of Sciences, Beijing
100190, China}

\author{Genfu Chen}
\affiliation{Beijing National Laboratory for Condensed Matter
Physics, Institute of Physics, Chinese Academy of Sciences, Beijing
100190, China}

\author{Shiliang Li}
\affiliation{Beijing National Laboratory for Condensed Matter
Physics, Institute of Physics, Chinese Academy of Sciences, Beijing
100190, China}
\affiliation{Collaborative Innovation Center of Quantum Matter, Beijing, China}

\author{Huiqian Luo}
\affiliation{Beijing National Laboratory for Condensed Matter
Physics, Institute of Physics, Chinese Academy of Sciences, Beijing
100190, China}

\author{Kristjan Haule}
\affiliation{Department of Physics, Rutgers University, Piscataway, NJ 08854, USA}

\author{Gabriel Kotliar}
\affiliation{Department of Physics, Rutgers University, Piscataway, NJ 08854, USA}
\affiliation{Brookhaven National Laboratory, Upton, NY 11973-5000, USA}

\author{Pengcheng Dai}
\email{pdai@rice.edu}
\affiliation{Department of Physics and Astronomy, Rice University, Houston, Texas 77005, USA}
\affiliation{Center for Advanced Quantum Studies and Department of Physics, Beijing Normal University, Beijing 100875, China}

\begin{abstract}
We use inelastic neutron scattering to study temperature and energy dependence of spin excitations
in optimally P-doped BaFe$_2$(As$_{0.7}$P$_{0.3}$)$_2$ superconductor ($T_c=30$ K) throughout the Brillouin zone.
In the undoped state, spin waves and paramagnetic spin excitations of BaFe$_2$As$_2$ stem from antiferromagnetic (AF) ordering
wave vector ${\bf Q}_{AF}=(\pm 1,0)$ and peaks near zone boundary at $(\pm1,\pm1)$ around 180 meV.  Replacing 30\% As by smaller
P to induce superconductivity, low-energy
spin excitations of BaFe$_2$(As$_{0.7}$P$_{0.3}$)$_2$
form a resonance in the superconducting state and high-energy spin excitations
now peaks around 220 meV
near $(\pm1,\pm1)$. These results are consistent with calculations from a
combined density functional theory and dynamical mean field
theory, and suggest that the decreased average
pnictogen height in BaFe$_2$(As$_{0.7}$P$_{0.3}$)$_2$ reduces
the strength of electron correlations
and increases the effective bandwidth of magnetic excitations.
\end{abstract}


\maketitle


Since the discovery of unconventional superconductivity in iron pnictides near antiferromagnetic (AF) instability  \cite{kamihara,cruz,SJiang,johnston,scalapino,Shibauchi,dai}, a central issue has been whether these materials are fundamentally different from copper oxide superconductors, where the magnetism and
superconductivity are derived from
Mott physics and its associated electron correlations \cite{palee,hirschfeld,chubukov,si,cfang,xuc,Basov11}.
Since iron pnictides have tetrahedrally coordinated nearest pnictogen atoms,
the $3d$ level in Fe ions splits into an $e_g$ state and a few hundred meV higher $t_{2g}$ state \cite{haule_NJP,zpyin11,Goerges,cclee,kruger,lv,ccchen,valenzeula}. Without Hund's rule interaction, the $e_g$ state would be
fully occupied with four of the six
Fe $3d$ electrons while the remaining two $3d$ electrons should reside in the $t_{2g}$ state crossing the Fermi level.
The presence of a strong Hund's coupling $J_H$, which tends to align spins of all the electrons on a given Fe-atom,
competes with the crystal field splitting and promotes high spin states of the Fe $3d$ electrons, resulting
in large charge and spin fluctuations.
Although it is generally accepted that electron correlations are also present in iron pnictides \cite{si,cfang,xuc,Basov11},
it remains unclear if the correlation strength is controlled by the on-site Hubbard $U$ interaction as in
the case of cuprates \cite{palee} or arises primarily from
the Hund's coupling $J_H$ within one atomic site \cite{haule_NJP,zpyin11,Goerges}.
The local moments formed by the Fe
$3d$ electrons, especially those in the $d_{xz}$, $d_{yz}$, and $d_{xy}$ orbitals, are coupled to their nearest neighbors by both the direct exchange associated with nearest neighbor Fe-Fe distance and
anisotropic superexchange interactions via hopping through the As pnictogen [Figs. 1(a), 1(b), and 1(c)].
By increasing the Fe-pnictogen distance, electron hopping between the nearest Fe ions becomes difficult, and the system
is localized with enhanced electron correlations. On the other hand, reducing the Fe-pnictogen distance facilitates the
electron hopping, and thus reducing the electron correlations.

Using density functional theory (DFT) combined with dynamical mean field theory (DMFT)
suitable for describing the Hund's coupling
in iron pnictides \cite{haule_NJP,zpyin11,Goerges,DMFT06,Haule_DMFT}, the evolution of spin excitations in electron and hole-doped BaFe$_2$As$_2$ can be calculated  \cite{hwpark,msliu,mwang}.  In particular,
the theory predicts that
spin-wave and spin excitation bandwidths, defined as the peak in energy dependence of the dynamic susceptibility
$\chi^{\prime\prime}(E)$ integrated near the AF zone boundary $(\pm1,\pm1)$ \cite{msliu,mwang},
in different iron pnictides are controlled by the iron-pnictogen distance and the valence of the Fe atoms similar to
electron correlations \cite{zpyin13}.  Experimentally, spin-waves and spin excitations in electron and hole-doped
$A$Fe$_2$As$_2$ ($A$=Ba, Sr, Ca) family of iron pnictides mapped out by inelastic neutron scattering experiments
throughout the Brillouin zone \cite{diallo,harriger11,ewings,jtpark12,HQLuo13} are qualitatively consistent with the DFT+DMFT calculations
and peak around 180 meV near $(1,1)$ \cite{msliu,mwang}.
Although neutron scattering experiments on spin waves of
NaFeAs, which has larger pnictogen height ($h_{Pn}=1.416$ \AA) \cite{slli} compared with that of BaFe$_2$As$_2$  [Fig. 1(b),
$h_{Pn}=1.36$ \AA], confirm the notion that increasing pnictogen height in iron pnictides decreases the spin-wave bandwidths
to $\sim$110 meV near $(1,1)$ and increases the electron correlation \cite{CLZhang14}, the crystal structures of NaFeAs and BaFe$_2$As$_2$ are rather different, and it is still unclear that varying the iron-pnictogen distance within one family of iron
pnictides can indeed control the electron correlations and spin excitation spectra.

\begin{figure} \includegraphics[scale=.45]{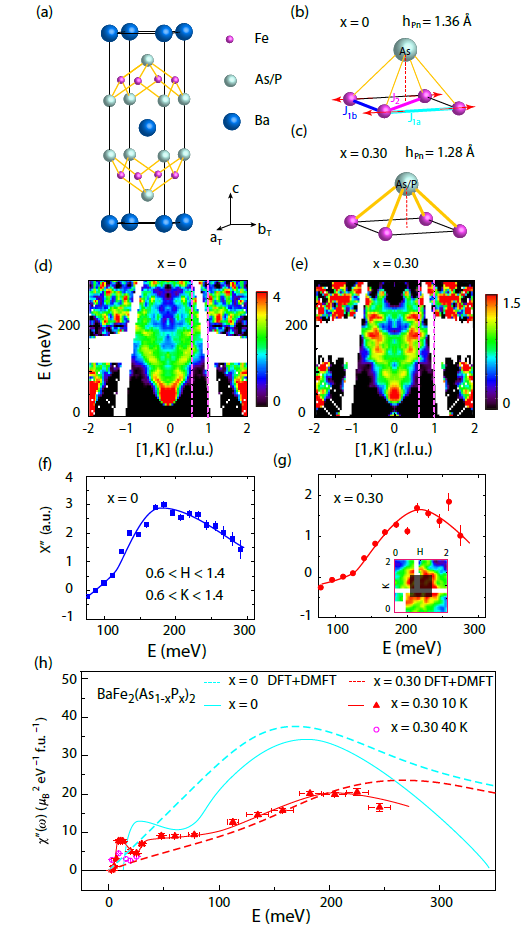}
\caption{(a) The crystal structure of BaFe$_2$(As$_{1-x}$P$_x$)$_2$. The purple, silvery and blue balls indicate Fe, As/P, and Ba positions, respectively. (b,c) Schematic diagrams of the FeAs tetrahedron, showing the average iron pnictogen height decreased from 1.36 \AA\ for BaFe$_2$As$_2$ to 1.28 \AA\ for BaFe$_2$(As$_{0.7}$P$_{0.3}$)$_2$ \cite{Allred2014}. (d)
The energy dependence of $S(Q,E)$ of
spin waves of BaFe$_2$As$_2$ along the $(1,K)$ direction
(with integration of $H$ from 0.9 to 1.1 rlu) after subtracting the background integrated
from $1.8<H<2.2$ and from $-0.25<K<0.25$ rlu with $E_i = 450$ meV at $T=10$ K
measured on MAPS \cite{harriger11}. (e) Identical projection
for spin excitations of BaFe$_2$(As$_{0.7}$P$_{0.3}$)$_2$ obtained on MAPS with $E_i = 450$ meV. The negative scattering below
$\sim$50 meV is due to errors in background subtraction.
Energy dependence of wave vector integrated [integration range is shown in the
shaded box in the inset of (g)] dynamic susceptibility $\chi^{\prime\prime}(E)$
for (f) BaFe$_2$As$_2$ and (g) BaFe$_2$(As$_{0.7}$P$_{0.3}$)$_2$.
The vertical dashed lines in (d) and (e) show wave vector integration range along the [0,K] direction.
 (h) Energy dependence of
the local dynamic spin susceptibility $\chi^{\prime\prime}(E)$
for BaFe$_2$As$_2$ (solid cyan line), BaFe$_2$(As$_{0.7}$P$_{0.3}$)$_2$
below (solid red triangle and solid red line) and above (open purple circles) $T_c$ with
corrected magnetic form factor.
Dashed cyan  and red lines are DFT+DMFT calculations for
 BaFe$_2$As$_2$ and BaFe$_2$(As$_{0.7}$P$_{0.3}$)$_2$, respectively.
}
\end{figure}

\begin{figure}[t] \includegraphics[scale=.45]{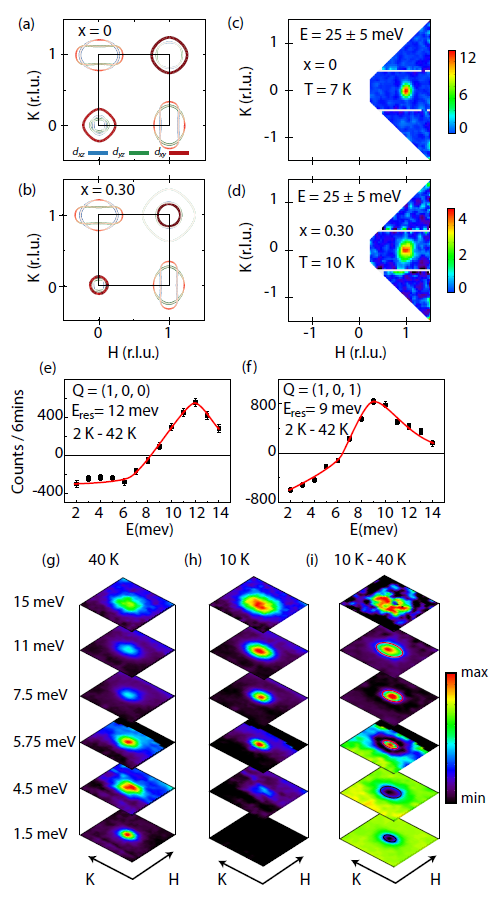}
\caption{
(a,b) Calculated Fermi surfaces of the $d_{xz}$, $d_{yz}$, and $d_{xy}$ orbitals
for BFe$_2$As$_2$ and BaFe$_2$(As$_{0.7}$P$_{0.3}$)$_2$, respectively.
(c,d) The corresponding  wave vector dependence of the low-energy ($E=25$ meV) spin excitations.
The color bars
represent the vanadium-normalized absolute spin wave intensity in
units of mbar/sr/meV/Fe. The temperature differences of
constant-wave vector scans at (e) ${\bf Q}=(1,0,0)$ and  (f) $(1,0,1)$ blow (2 K) and
above (42 K) $T_c$ obtained using the EIGER triple-axis spectrometer.
The modulation of resonance (superconductivity-induced intensity gain) with $L$ is consistent with previous
experiment \cite{CHLee2013}.
Constant energy slices of the spin excitations as a function of
increasing energy (g) at 40 K and (h) 10 K.  (i) Temperature difference between 10 K and 40 K, showing clearly intensity gain in the energy region of 8 meV to
11 meV.
 }
\end{figure}

In this paper, we present inelastic neutron scattering
studies of temperature and energy dependence of spin excitations in BaFe$_2$(As$_{0.7}$P$_{0.3}$)$_2$
superconductor ($T_c=30$ K) \cite{SJiang,Shibauchi}.
We chose BaFe$_2$(As$_{0.7}$P$_{0.3}$)$_2$ because it is
the optimal isovalently doped BaFe$_2$As$_2$. At the same time, it has the average pnictogen height ($h_{Pn}=1.28$
\AA) significantly smaller than that of BaFe$_2$As$_2$ due to the smaller size of the P dopants [Figs. 1(b) and 1(c)] \cite{johnston}.
Since the average pnictogen height of
BaFe$_2$(As$_{1-x}$P$_{x}$)$_2$ decreases continuously from BaFe$_2$As$_2$ to
BaFe$_2$(As$_{0.7}$P$_{0.3}$)$_2$ without modifying much
the in-plane Fe-Fe distance or changing the valence of Fe
\cite{johnston}, we expect weaker electron correlations and wider spin excitation bandwidth in
BaFe$_2$(As$_{0.7}$P$_{0.3}$)$_2$ compared with that of BaFe$_2$As$_2$ \cite{zpyin13}.

Figures 1(d) and 1(e) show the energy dependence of
$S(Q,E)$ of spin excitations for BaFe$_2$As$_2$ and BaFe$_2$(As$_{0.7}$P$_{0.3}$)$_2$,
respectively.  While spin-wave dispersions of BaFe$_2$As$_2$ reach zone boundary positions $(1,\pm 1)$
around 200 meV [Fig. 1(d)], dispersions of spin excitations of  BaFe$_2$(As$_{0.7}$P$_{0.3}$)$_2$ become steeper, and reach
$(1,\pm 1)$ at energies well above 200 meV [Fig. 1(e)].
Indeed, we find that spin excitations in BaFe$_2$(As$_{0.7}$P$_{0.3}$)$_2$ have lower intensity but
 larger energy
bandwidth than that of BaFe$_2$As$_2$ \cite{harriger11}, as revealed in
 the energy dependence of the dynamic susceptibility $\chi^{\prime\prime}(E)$
integrated over the dashed vertical lines in Figs. 1(d) and 1(e) ($0.6\leq H\le 1.4$ and $0.6\leq K\le 1.4$)
near the zone boundary $(1,\pm 1)$ [Figs. 1(f) and 1(g)].
These results are also consistent with
the energy
dependence of the local dynamic susceptibility and DFT+DMFT calculations of $S(Q,E)$ [Fig. 1(h), Figs. 2-4].
Therefore, the decreased average pnictogen height in BaFe$_2$(As$_{1-x}$P$_{x}$)$_2$ decreases
the electron correlations and increases the overall spin excitation energy bandwidths.

Our neutron scattering experiments were carried out on the MAPS and 4SEASONS chopper spectrometers at ISIS, Rutherford Appleton Laboratory, UK and Japan Proton Accelerator Research Complex, Japan, respectively.
Some measurements are also carried out on EIGER triple-axis spectrometer in Paul Scherrer Institut, Switzerland.
Our BaFe$_2$(As$_{0.7}$P$_{0.3}$)$_2$
single crystals were grown by a self-flux method \cite{DHu2015}.  For P-doped BaFe$_2$(As$_{1-x}$P$_{x}$)$_2$
near $x=0.3$, the collinear static AF order in BaFe$_2$As$_2$ is suppressed and superconductivity reaches optimal value
at $T_c=30$ K \cite{Allred2014}.   We coaligned $\sim$17 g of single
crystals in the $[H,H,L]$ scattering plane with a mosaic
$<$7$^\circ$. To facilitate easy comparison with spin waves in BaFe$_2$As$_2$, which has an orthorhombic AF ground state \cite{harriger11},
we define the wave vector ${\bf Q}$ at ($q_x$,$q_y$,$q_z$) in
\AA$^{-1}$ as (\textit{H},\textit{K},\textit{L}) =
($q_xa/2\pi$,$q_yb/2\pi$,$q_zc/2\pi$) where $a=
b\approx 5.6$ \AA\ and $c = 12.87$ \AA\ using the orthorhombic magnetic
unit cell notation where low-energy
spin excitations are expected to stem from the in-plane wave vector positions ${\bf Q}_{AF}=(\pm 1,0)$ and $(0,\pm 1)$.
For chopper spectrometer inelastic neutron scattering measurements, the
incident beam energies were $E_i$ = 35, 80, 250, 450 meV at MAPS and
$E_i$ = 13, 21, 82, 313 meV at 4SEASONS
with $k_i$
parallel to the $c$-axis. Spin excitation intensity was normalized to
absolute units using a vanadium standard ($\sim$ 30\% error).

Since there is no evidence that the P dopant forms long range order in
BaFe$_2$(As$_{1-x}$P$_{x}$)$_2$, we use an effective pnictide position in the DFT+DMFT calculations to simulate the physical consequence of P doping. In BaFe$_2$(As$_{0.71}$P$_{0.29}$)$_2$,
the As and P heights are 1.332 and 1.151 \AA\ from the Fe plane, respectively \cite{Rotter10}.
We therefore take an effective pnicogen height of 1.28 \AA\ in our calculation, which is the average height of As and P
in BaFe$_2$(As$_{0.71}$P$_{0.29}$)$_2$ determined experimentally. This effective As/P height (1.28 \AA) is less than the As height (1.36 \AA) in the BaFe$_2$As$_2$ but substantially larger than the P height (1.19 \AA) in the BaFe$_2$P$_2$.

In previous inelastic neutron scattering studies of low-energy
spin excitations in powder \cite{Ishikado} and single crystals \cite{CHLee2013} of
optimally P-doped BaFe$_2$(As$_{1-x}$P$_{x}$)$_2$, a neutron spin resonance coupled to superconductivity has been
identified similar to other iron based superconductors \cite{Lumsden2009,SChi2009,DSInosov2010}.
A key conclusion of the work is that the energy of the resonance in
BaFe$_2$(As$_{0.63}$P$_{0.34}$)$_2$
is dispersive along the $c$ axis, indicating its close connection to the three
dimensional AF spin correlations \cite{CHLee2013}.  In hole and electron-doped BaFe$_2$As$_2$, the
wave vector evolution of the low-energy
spin excitations and the resonance can be well described by
quasiparticle excitations through doping dependent hole and electron Fermi surfaces \cite{mwang,HQLuo2012}.
Although substituting P for As in BaFe$_2$(As$_{1-x}$P$_{x}$)$_2$ is expected to be isovalent, angle resolved photoemission
spectroscopy (ARPES) experiments reveal changed hole and electron Fermi surfaces from BaFe$_2$As$_2$
to BaFe$_2$(As$_{0.7}$P$_{0.3}$)$_2$ \cite{YZhang12}. DFT+DMFT calculations also show that with increasing P-doping, the hole Fermi surface with dominating $d_{xy}$ orbital character shrinks whereas the hole Fermi surfaces with dominating $d_{xz}$ and $d_{yz}$ ($d_{z^2}$ near $k_z=\pi$) orbital characters expand and become more three-dimensional along the $k_z$ direction [Figs. 2(a) and 2(b)]. However, the electron Fermi surfaces do not change significantly. Therefore, the electron-hole Fermi surface nesting condition becomes worse with P-doping. This is consistent with the resulting changes in wave vector dependence of spin excitations [Figs. 2(c) and 2(d)], where the low energy spin excitations become weaker and more diffusive in the momentum space in BaFe$_2$(As$_{0.7}$P$_{0.3}$)$_2$.  Figure 2(e) and 2(f) shows constant-${\bf Q}$ scans at ${\bf Q}=(1,0,0)$ and $(1,0,1)$, respectively, below and above
$T_c$.  Consistent with previous work \cite{CHLee2013}, we find that superconductivity-induced resonance is clearly dispersive, occurring at $E_{res}=12$ meV at ${\bf Q}=(1,0,0)$ and $E_{res}=9$ meV at ${\bf Q}=(1,0,1)$.
Figure 2(g) and 2(h) summarizes energy dependence of the low-energy spin excitations at 40 K ($T \approx T_c+10$ K) and
10 K ($T \approx T_c-20$ K), respectively.  Given the dispersive nature of the resonance, neutron time-of-fight measurements with a fixed incident energy
and fixed sample rotation angle in Figs. 2(g) and 2(h) will probe a region of the excitation energies with different $L$ values.
Figure 2(i) is temperature differences plot, revealing a clear neutron spin resonance  in the energy region of
$E\approx 11$ meV and a spin gap below the resonance energy.

\begin{figure}[t] \includegraphics[scale=.25]{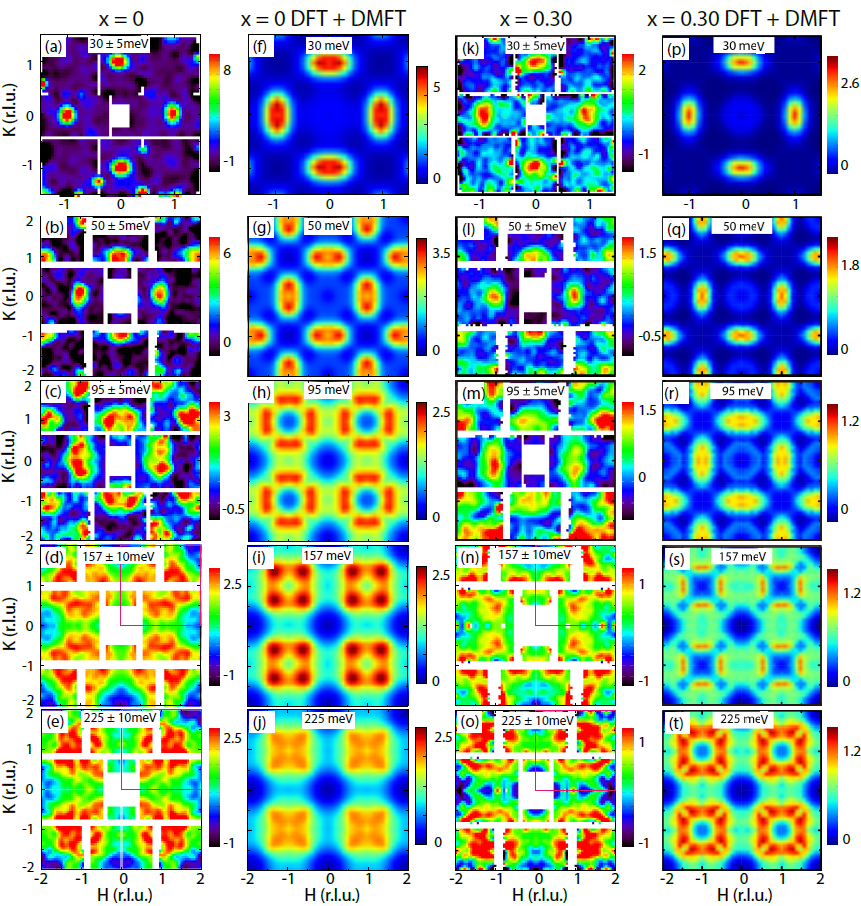}
\caption{Wave vector dependence of spin waves of BaFe$_2$As$_2$ at 7 K and spin excitations of
BaFe$_2$(As$_{0.7}$P$_{0.3}$)$_2$ at 10 K for
energy transfers of (a,k) \textit{E} = 30$\pm 5$ meV [${E}_i$ =
80 meV and ${\bf q} = (H,K,3)$]; (b,i) $E =
50\pm 5$ meV [$\textit{E}_i = 250$ meV and ${\bf q} =
(H,K,3)$]; (c,m) $E = 95\pm 5$ meV
[$E_i = 250$ meV and ${\bf q} = (H,K,5)$]; (d,n)
$E = 157\pm 10$ meV [$\textit{E}_i = 450$ meV and ${\bf q} =
(\textit{H},\textit{K},6)$]; (e,o) $\textit{E} = 225\pm 10$ meV
[$\textit{E}_i = 450$ meV and ${\bf q} = (\textit{H},\textit{K},9)$].
In all cases, the $\pm$ meV indicates
the energy integration range.  The red boxes in (d,e,n,o) indicate
 regions that contain nonduplicate data from four fold symmetrizing of
the raw data (meaning only the data within the red box is statistically significant, and data in other region of reciprocal space are mirror images of the red box data).
(f-j) and (p-t) Calculations of
identical energy slices from DFT+DMFT method \cite{supplementary}.
}
\end{figure}

Assuming that isovalent P-doping in BaFe$_2$(As$_{1-x}$P$_{x}$)$_2$ does not change the valence of Fe,
the total moment sum rule requires the total magnetic spectral weight $M_0$, when integrated over all energy
and momentum space [$M_0^2=M^2+\left\langle{\bf m}^2\right\rangle=g^2S(S+1)$, where $M$ is the static ordered
moment, $\left\langle{\bf m}^2\right\rangle$ is the local fluctuating moment, $g\approx 2$ is the
Land$\rm \acute{e}$ factor, and $S$ is the spin.], to be independent of P-doping \cite{dai}.
From Figs. 2(c) and 2(d), we see reduced low-energy
spin excitation spectral weight in BaFe$_2$(As$_{0.7}$P$_{0.3}$)$_2$ compared with
BaFe$_2$As$_2$, as revealed by the DFT+DMFT calculation [Figs. 1(h) and 4].

Figures 3(a)-3(e) and 3(k)-3(o) compare the two-dimensional
constant-energy ($E$) images of $S(Q,E)$ of spin waves of BaFe$_2$As$_2$ \cite{harriger11} and
spin excitations of BaFe$_2$(As$_{0.7}$P$_{0.3}$)$_2$
in the $(H,K)$ scattering plane at different energies.
Figures 3(a)-3(e) show the evolution of
spin waves of BaFe$_2$As$_2$ at energy transfers of
$E=30\pm 5$, $50\pm 5$, $95\pm 5$, $157\pm 10$, and $225\pm 10$ meV, respectively.
The corresponding spin excitations of BaFe$_2$(As$_{0.7}$P$_{0.3}$)$_2$ are shown in Figs. 3(k)-3(o).
At $E=30\pm 5$ [Figs. 3(a) and 3(k)] and $E=50\pm 5$ meV [Figs. 3(b) and 3(l)], spin excitations in
BaFe$_2$(As$_{0.7}$P$_{0.3}$)$_2$
form transversely elongated ellipses centered at the in-plane AF zone centers
$(\pm 1,0)$ and $(0,\pm 1)$ of the undoped BaFe$_2$As$_2$, but with considerably lower intensity.
On increasing the energies to $E=95\pm 5$ meV [Fig. 3(m)], spin
excitations of BaFe$_2$(As$_{0.7}$P$_{0.3}$)$_2$ begin to
split transversely from $(\pm 1,0)$, similar to that of
spin waves of BaFe$_2$As$_2$ [Fig. 3(c)].
On further increasing the energy to
$E=157\pm 10$ [Fig. 3(n)] and $E=225\pm 10$ meV [Fig. 3(o)], spin excitations
of BaFe$_2$(As$_{0.7}$P$_{0.3}$)$_2$
form anisotropic rings centered around
 $(\pm 1,\pm 1)$.  Figure 1(d) and 1(e) shows spin waves
 of BaFe$_2$As$_2$ at $E=157\pm 10$ and $E=225\pm 10$ meV, respectively.
We see that spin waves of BaFe$_2$As$_2$ at $E=225\pm 10$ meV nearly form a solid spot
at $(\pm 1,\pm 1)$, suggesting that the system has already reached zone boundary at this energy.
For comparison, spin excitations of BaFe$_2$(As$_{0.7}$P$_{0.3}$)$_2$ at
$E=225\pm 10$ meV still have a ring structure near $(\pm 1,\pm 1)$ [Fig. 3(o)],
as confirmed by comparison of constant-energy cuts across the data (SFig. 5 in \cite{supplementary}).
These results suggest
a higher zone boundary energy for BaFe$_2$(As$_{0.7}$P$_{0.3}$)$_2$.

\begin{figure}[t] \includegraphics[scale=.3]{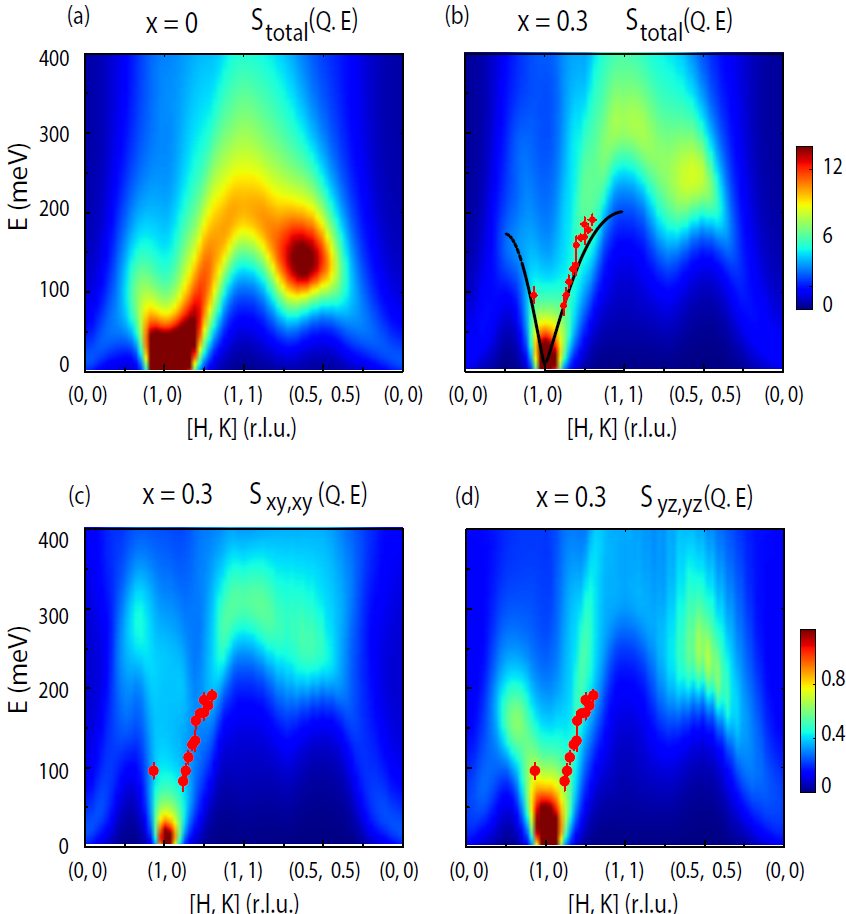}
\caption{ (a)
Calculated total dynamic magnetic structure factor $S(Q,E)$ for (a) BaFe$_2$As$_2$ and (b) BaFe$_2$(As$_{0.7}$P$_{0.3}$)$_2$ using
DFT+DMFT. The solid red points in (b) are data from cuts to Fig. 3 and the solid line is the dispersion of BaFe$_2$As$_2$ from
\cite{harriger11}.
(c,d) Calculated dynamic magnetic structure factors from the
$d_{xy}$-$d_{xy}$ and $d_{yz}$-d$_{yz}$ intra-orbital contributions, respectively.
}
\end{figure}

To quantitatively compare the experimental results with
a combined DFT+DMFT theory \cite{zpyin13}, we show in Figs. 3(f)-3(j) and
3(p)-3(t) calculated
wave vector dependence of
spin excitations of BaFe$_2$As$_2$ and
 BaFe$_2$(As$_{0.7}$P$_{0.3}$)$_2$, respectively, at energies in Fig. 3(a)-3(e) \cite{zpyin13}.
We see that
spin excitations at different energies
obtained from the DFT+DMFT calculation in  Fig. 3(p)-3(t) have much similarities
with the experimental data [Figs.~3(k)-3(o)].

Figure 4(a) and 4(b) shows the calculated dynamical magnetic structure factor $S(Q,E)$ for
BaFe$_2$As$_2$ and
BaFe$_2$(As$_{0.7}$P$_{0.3}$)$_2$, respectively. Our calculation reveals considerable
magnetic spectral weight for energies above 300 meV for both samples contrasting to the vanishing
local dynamic susceptibility for energies above 300 meV in BaFe$_2$As$_2$ [Fig. 1(h)] \cite{harriger12}.
 The experimentally determined spin excitation dispersion along the $[1,K]$ direction
for BaFe$_2$(As$_{0.7}$P$_{0.3}$)$_2$
is well captured by the DFT+DMFT calculations [Fig. 4(b)]. For comparison, we also plot in Fig. 4(b) the experimentally determined dispersion along the same direction for BaFe$_2$As$_2$ as solid line \cite{harriger11}.
To understand different orbital contributions to the spin excitations, we show in Figs. 4(c) and (d)
 the $d_{xy}$-$d_{xy}$ and $d_{yz}$-d$_{yz}$ intra-orbital contribution to the dynamic susceptibility.
We see that low-energy spin excitations near $(1,0)$ are mostly contributed by excitations involving the $d_{yz}$ orbital,
while high-energy spin excitations around 300 meV near $(1,1)$ come mostly from excitations related to $d_{xy}$ orbital.
This is in stark contrast with spin excitations in Co-doped LiFeAs compound, where the low-energy spin excitations are dominated by contributions from the $d_{xy}$ orbital \cite{Yu}. In Ref. \cite{Yu}, it was concluded that the $d_{xz/yz}$ orbitals play an important role in the superconductivity of LiFeAs since their absence in
LiFe$_{0.88}$Co$_{0.12}$As
suppresses superconductivity. The strong low-energy spin excitations contributed by the $d_{yz/xz}$ orbitals
in BaFe$_2$(As$_{0.7}$P$_{0.3}$)$_2$ shown in Fig. 4(d)
suggest that nesting of the $d_{yz/xz}$ orbitals
are good for high temperature superconductivity.

Figure 1(h) compares the energy dependence of the local dynamic spin susceptibility for
BaFe$_2$As$_2$, BaFe$_2$(As$_{0.7}$P$_{0.3}$)$_2$, and DFT+DMFT calculated values.
We see that the peak for local dynamic spin susceptibility for BaFe$_2$(As$_{0.7}$P$_{0.3}$)$_2$ occurs around 220 meV,
while it is around 180 meV for BaFe$_2$As$_2$.  These results are consistent with energy cuts
near the zone boundary for these materials shown in Figs. 1(f) and 1(g).
 The total
fluctuating magnetic moments for BaFe$_2$(As$_{0.7}$P$_{0.3}$)$_2$ and BaFe$_2$As$_2$ are
$\left\langle m^2\right\rangle \approx 1.6\pm 0.2$ (below $\sim$250 meV) and 3.6 $\mu_B^2$ per
Fe, respectively \cite{harriger12}.
This means that the fluctuating moments of BaFe$_2$(As$_{0.7}$P$_{0.3}$)$_2$
are smaller than that of BaFe$_2$As$_2$ within our energy integration region, consistent with the presence of more magnetic spectral weight at higher energies
or a reduced fluctuating moment.  These results thus suggest that the decreased iron pnictogen height in iron
pnictides from BaFe$_2$As$_2$ to BaFe$_2$(As$_{0.7}$P$_{0.3}$)$_2$ increases
the spin excitation bandwidth
and decreases the electron correlation effects. Since the reduced pnicogen height due to P doping increases the indirect hopping between Fe $3d$ orbitals and pnicogen $p$ orbitals and weakens the kinetic frustration as the direct hopping between Fe $3d$ orbitals remains almost the same as the Fe-Fe distance changes negligible with P-doping, the band widths of the Fe $3d$ orbitals increase with increasing P doping \cite{supplementary} and the electronic correlation effects decrease. The increased band widths lead to a reduction of the one-particle Green's function, and thus a reduction in the bare two-particle susceptibility. Neglecting the change in the two-particle vertex function due to P doping, this reduction in the bare two-particle susceptibility is responsible for the reduced low energy spin excitation intensity. Similarly, the spin excitation band widths increases due to the reduced pnicogen height in BaFe$_2$(As$_{0.7}$P$_{0.3}$)$_2$.

The reduction of the pnicogen height in BaFe$_2$(As$_{0.7}$P$_{0.3}$)$_2$ from BaFe$_2$As$_2$ reduces
the low energy spin excitation intensity centered at $Q_{AF}$ and eliminates the static long-range AF order in the undoped
BaFe$_2$As$_2$. At the same time, the pnicogen height in BaFe$_2$(As$_{0.7}$P$_{0.3}$)$_2$ is still high enough to maintain intermediate electronic correlation strength with sufficient low-energy spin fluctuations to mediate superconductivity.
For spin excitation mediated superconductors
\cite{scalapino}, superconductivity is controlled by the
effective magnetic exchange coupling $J$ and the strength of
electron-spin excitation coupling \cite{mwang}.  Since the effective magnetic
exchange couplings in
BaFe$_2$(As$_{0.7}$P$_{0.3}$)$_2$
 are considerably larger than those of
the BaFe$_2$As$_2$, it would be interesting to compare superconductivity induced changes in
spin excitations of BaFe$_2$(As$_{0.7}$P$_{0.3}$)$_2$ and
electron/hole-doped BaFe$_2$As$_2$ \cite{mwang}.  By comparing the absolute intensity changes of the
resonance below and above $T_c$, we find that spin excitations changes across $T_c$ are still much larger than the
superconducting condensation energy \cite{supplementary,ZDiao}, thus supporting the notion that magnetism is crucial for the
superconductivity of  BaFe$_2$(As$_{0.7}$P$_{0.3}$)$_2$.

In summary, we have used inelastic neutron scattering to map out spin excitations of
isovalently doped BaFe$_2$(As$_{0.7}$P$_{0.3}$)$_2$.
By comparing spin excitations of this material with those of BaFe$_2$As$_2$ and DFT+DMFT calculations, we conclude that iron pnictogen height in iron pnictides directly controls the spin excitation bandwidth and electron correlations.  These results are consistent with the idea that
electron correlations in iron-based superconductors arise primarily from
the Hund's coupling $J_H$, and low-energy spin excitations are consequences of nesting between hole and electron Fermi surfaces.

The work at IOP, CAS is supported by NSFC (Projects No. 11374011 and
No. 11374346), MOST of
China (973 projects: 2012CB821400 and 2015CB921302),
and The Strategic Priority Research Program (B) of the
Chinese Academy of Sciences (Grant No. XDB07020300).
Neutron scattering work at Rice is supported by the U.S. DOE, Office of Basic Energy Sciences, under
Contract No. DE-SC0012311. Part of the
materials work at Rice University is supported by the Robert
A. Welch Foundation Grant No. C-1839.
The neutron experiment at the Materials and Life Science Experimental Facility of J-PARC was performed under an user program (Proposal No. 2014B0277).
The computational work at Rice and Rutgers are supported by NSF DMREF DMR-1436006 and DMR-1435918, respectively.

%\bibliography{NoEndingPoint}

\end{document}